\documentclass[pdflatex,sn-mathphys-num, iicol]{sn-jnl}
\usepackage{graphicx}%
\usepackage{multirow}%
\usepackage{amsmath,amssymb,amsfonts}%
\usepackage{amsthm}%
\usepackage{mathrsfs}%
\usepackage[title]{appendix}%
\usepackage{xcolor}%
\usepackage{textcomp}%
\usepackage{manyfoot}%
\usepackage{booktabs}%
\usepackage{algorithm}%
\usepackage{algorithmicx}%
\usepackage{algpseudocode}%
\usepackage{listings}%
\usepackage{capt-of}

\theoremstyle{thmstyleone}%
\theoremstyle{thmstyletwo}%

\theoremstyle{thmstylethree}%

\begin{document}

\title[Article Title]{4DMulti: automated multicomponent identification at complex material interfaces}

\author[1,2]{\fnm{Haoran} \sur{Zhang}}
\email{zhr20040823@sjtu.edu.cn}
\equalcont{These authors contributed equally to this work.}

\author[1,2]{\fnm{Zian} \sur{Mao}}
\email{asakura\_kukii@sjtu.edu.cn}
\equalcont{These authors contributed equally to this work.}

\author[1]{\fnm{Shufen} \sur{Chu}}
\email{shfchu@sjtu.edu.cn}

\author[1,3]{\fnm{Xiaoya} \sur{He}}
\email{hexiaoya@sjtu.edu.cn}

\author[1]{\fnm{Yuyan} \sur{Guan}}
\email{Gyy0608@sjtu.edu.cn}

\author[1,2]{\fnm{Antong} \sur{Yang}}
\email{snowynight@sjtu.edu.cn}

\author[1]{\fnm{Mingze} \sur{Li}}
\email{mingze.l@sjtu.edu.cn}

\author*[3]{\fnm{Xiaoqin} \sur{Zeng}}
\email{xqzeng@sjtu.edu.cn}

\author*[1]{\fnm{Yujun} \sur{Xie}}
\email{yujun.xie@sjtu.edu.cn}

\affil[1]{
    \orgdiv{Global Institute of Future Technology},
    \orgname{Shanghai Jiao Tong University},
    \postcode{200240},
    \state{Shanghai},
    \country{China}
}

\affil[2]{
    \orgdiv{Global College},
    \orgname{Shanghai Jiao Tong University},
    \postcode{200240},
    \state{Shanghai},
    \country{China}
}

\affil[3]{
    \orgdiv{School of Materials Science and Engineering},
    \orgname{Shanghai Jiao Tong University},
    \postcode{200240},
    \state{Shanghai},
    \country{China}
}


\abstract{Mapping crystalline phases at heterogeneous interfaces is essential for understanding material performance and degradation. However, structural heterogeneity, phase overlap, and local disorder complicate diffraction interpretation, while growing data volumes make manual analysis increasingly impractical. We introduce 4DMulti, a physics-guided learning framework for automated multicomponent identification from large-scale four-dimensional scanning transmission electron microscopy (4D-STEM) data. The supporting diffraction data resource comprises over 6 million high-quality experimental patterns and labeled patterns generated by Sim2real. A retrieval-conditioned latent diffusion transformer (Sim2real) translates simulated patterns into experimental-style examples under constraints designed to preserve Bragg geometry, while a rotation-invariant coordinate convolutional network identifies phases across in-plane rotations. 4DMulti achieves 98.82\% classification accuracy on a five-phase experimental nanoparticle benchmark, with ablation studies supporting the complementary benefits of domain adaptation and rotation-invariant classification. We define diffraction-inferred structural complexity (DISC), a normalized predictive entropy score that quantifies phase-assignment ambiguity within a specified candidate phase library. We apply 4DMulti to generate structural maps of superconducting heterostructures, corroded alloy surfaces, and degraded solid-state battery interfaces down to single-nanometer spatial resolution. 4DMulti connects simulation-derived crystallographic knowledge to automated experimental interpretation, establishing a foundation for scalable analysis of complex interfaces and data-driven discovery of interfacial design principles.}

\keywords{Four-dimensional scanning transmission electron microscopy,
Simulation-to-experiment domain adaptation,
Rotation-invariant neural networks,
Predictive uncertainty}

\onecolumn
\maketitle

\twocolumn
\section*{Introduction}

Identifying the constituent phases at heterogeneous material interfaces is essential for linking local structure to macroscopic performance and degradation\cite{ref01,ref02}. For example, buried interfaces influence charge transport and stability in semiconductor devices\cite{ref03,ref04,ref05}, while nanoscale chemical and microstructural heterogeneity shapes localized corrosion and crack initiation in structural alloys\cite{ref06}. However, these interfaces often contain multiple crystalline phases and nanoscale disordered regions, with overlapping domains further complicating structural identification\cite{ref01,ref02}. Resolving this complexity requires characterization approaches that combine high spatial resolution and structural sensitivity with reliable automated interpretation, supporting both fundamental research and industrial applications\cite{ref01,ref02,ref03}.

Conventional bulk X-ray and neutron scattering provide detailed structural information averaged over large volumes, but generally cannot directly map nanoscale heterogeneity\cite{ref07,ref08}. High-resolution transmission electron microscopy (HRTEM) can resolve atomic-scale features, although obtaining statistically representative structural information across extended areas remains demanding\cite{ref09}. These complementary capabilities motivate approaches that connect local structural measurements to their spatial distribution across heterogeneous interfaces\cite{ref10,ref11}.

\begin{figure*}[!htp]
  \centering
  \includegraphics[width=\textwidth]{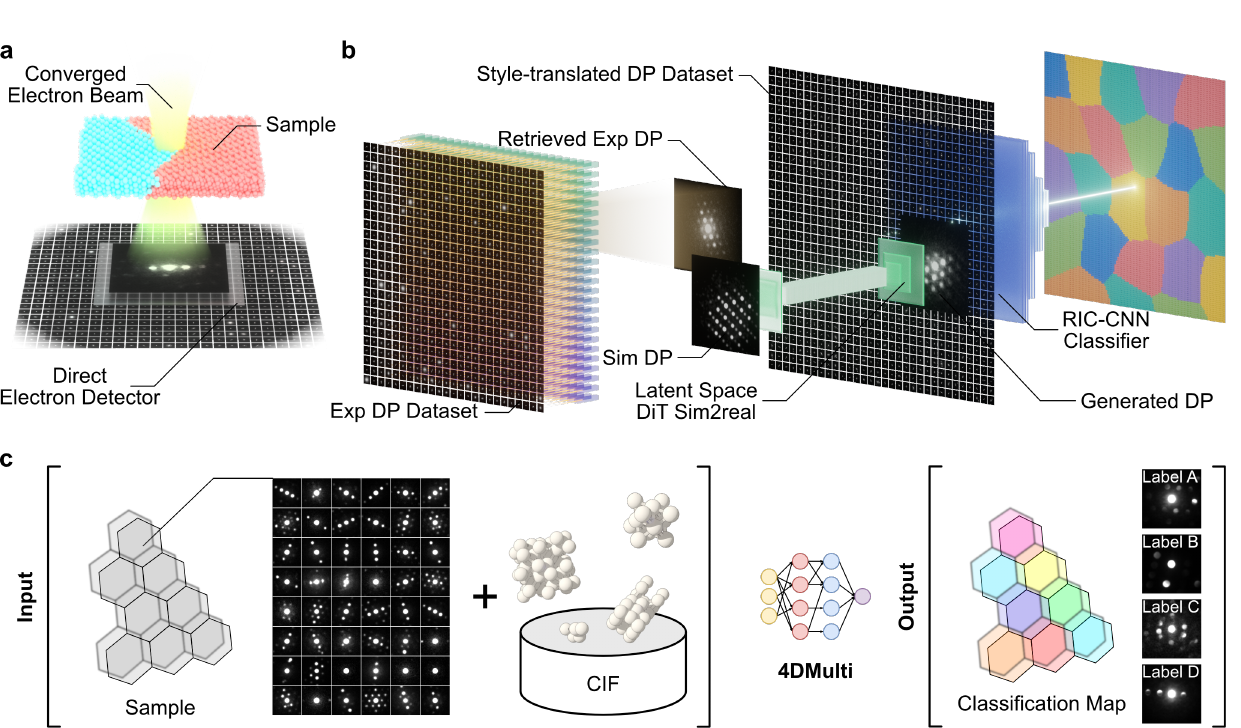}
\caption{\textbf{Overview of the 4DMulti framework for 4D-STEM phase and component analysis.}
\textbf{a}, Schematic of 4D-STEM data acquisition showing the converged electron beam scanning the sample and recording DPs onto a direct electron detector.
\textbf{b}, Architecture of the 4DMulti framework. Simulated DPs (Sim DP) derived from candidate crystal structures and experimental reference patterns (Retrieved Exp DP) from the experimental dataset (Exp DP Dataset) are fed into a latent-space Diffusion Transformer (DiT) Sim2real module. This module performs domain/style translation to generate realistic DPs (Generated DP), constructing a large-scale style-translated DP dataset. A rotation-invariant coordinate convolutional neural network (RIC-CNN) classifier is then trained on this dataset to perform pixel-level classification across the scanned area.
\textbf{c}, Inputs and outputs of 4DMulti. The framework takes experimental 4D-STEM scan datasets and crystallographic information files (CIFs) of candidate crystal structures as inputs, processes them through 4DMulti, and outputs a spatially resolved multicomponent classification map along with phase/grain identification labels. The illustrated output classes are schematic examples and do not restrict the number of candidate components.}
\label{fig:4DMulti_overview}
\end{figure*}

Four-dimensional scanning transmission electron microscopy (4D-STEM) provides this connection by recording a two-dimensional diffraction pattern (DP) at each position of a raster-scanned electron probe (Fig.~\ref{fig:4DMulti_overview}a)\cite{ref10,ref11}. The resulting dataset combines two real-space and two reciprocal-space dimensions, encoding local crystallinity, phase distributions, and structural defects\cite{ref12,ref13,ref14}. Direct electron detectors enable rapid acquisition and low-dose measurements across extended fields of view, with spatial resolution governed by the probe and specimen conditions and sampling determined by the scan step\cite{ref10,ref14}. At kilohertz to tens-of-kilohertz acquisition rates, individual scans can contain millions of DPs and generate tens to hundreds of gigabytes of data\cite{ref15,ref16,ref17}. Converting these measurements into reliable structural maps remains a central challenge despite advances in automated acquisition, preprocessing, and analysis\cite{ref15,ref16,liu2026preprocessing}. The central bottleneck is therefore not simply handling data volume, but converting large numbers of DPs into crystallographically meaningful assignments without repeated pattern-by-pattern expert interpretation.

Unsupervised machine-learning approaches, including non-negative matrix factorization and hierarchical clustering, have enabled segmentation of large 4D-STEM datasets into spatially coherent domains\cite{ref18,ref19,ref20}. However, mathematical similarity between patterns does not uniquely establish crystallographic identity: assigning phase labels generally requires subsequent interpretation or comparison with reference diffraction data\cite{ref19,ref20,ref22}. Results can also depend on preprocessing, dimensionality reduction, and user-defined parameters such as the number of clusters\cite{ref20,ref22}. These limitations become particularly important at interfaces, where diffraction signals from different phases may overlap or appear similar, making it difficult to assign a unique phase identity to each cluster.

A route towards automated interpretation is to encode candidate crystal structures in extensible libraries of simulated DPs, with labels supplied by the generating structures rather than manual annotation of experimental data. Supervised diffraction analysis has enabled structure factor recovery for strain mapping\cite{ref25} and orientation-agnostic classification of crystal systems and space groups\cite{ref29}. However, extending these methods to phase mapping at heterogeneous interfaces presents three challenges\cite{ref24}. First, differences in specimen thickness, structural disorder, detector response, and noise can limit transfer from simulated training data to experimental patterns\cite{ref24,ref25,ref26,ref27}. Second, phase predictions should remain consistent under in-plane rotations, a property not inherently guaranteed by standard convolutional neural networks (CNNs)\cite{ref28,ref29}. Third, a single experimental pattern can contain contributions from multiple phases along the electron-beam direction, making a single-class description incomplete or ambiguous. A useful framework must therefore connect simulated structural labels to experimental diffraction features while assessing how decisively each pattern can be assigned within the candidate phase library.

Here, we introduce 4DMulti for automated multicomponent identification and nanoscale phase mapping at complex interfaces from large-scale 4D-STEM data. Our diffraction data resource comprises over 6 million high-quality experimental patterns and over 200,000 labeled, experimental-style patterns generated by Sim2real. This resource supports a unified workflow linking candidate crystal structures to experimental observations. A retrieval-conditioned latent Diffusion Transformer (DiT)\cite{ref30,ref31} translates simulated patterns into experimental-style training data under constraints designed to preserve Bragg geometry. A rotation-invariant coordinate CNN (RIC-CNN) identifies candidate phases across in-plane rotations without rotational data augmentation\cite{ref32}. Given experimental diffraction data and a specified candidate phase library, 4DMulti automates phase assignment across the scanned region without manual interpretation of individual patterns. We define diffraction-inferred structural complexity (DISC) as normalized predictive entropy to map assignment ambiguity within the candidate phase library. We evaluate 4DMulti on an experimental nanoparticle benchmark and three interfacial systems: a superconducting heterostructure, a corroded magnesium alloy, and a moisture-degraded sulfide solid electrolyte. Together, these capabilities provide a route towards fully automated, library-guided crystallographic analysis, shifting the focus from interpreting individual DPs to systematically mapping the phase distributions and assignment ambiguity of complex material interfaces.

\section*{Results}

\subsection*{Overview of 4DMulti}

4DMulti is an end-to-end multi-component classification framework for 4D-STEM data, as summarized in Fig.~\ref{fig:4DMulti_overview}b. The overall architecture takes as input an unlabeled experimental diffraction dataset and a candidate set of crystal structures represented by crystallographic information files (CIFs) (see Supplementary Table~1). These candidate structures are enumerated from prior knowledge, complementary characterization, or physically motivated hypotheses regarding the interface composition. Multi-slice simulations are then performed over random crystal orientations to generate a labeled diffraction dataset (see Supplementary Fig.~1). By integrating experimental measurements with the simulation-derived prior, 4DMulti assigns each DP to one candidate component and thereby produces a spatially resolved component map of the interface.

\begin{figure*}[!htp]
  \centering
  \includegraphics[width=\textwidth]{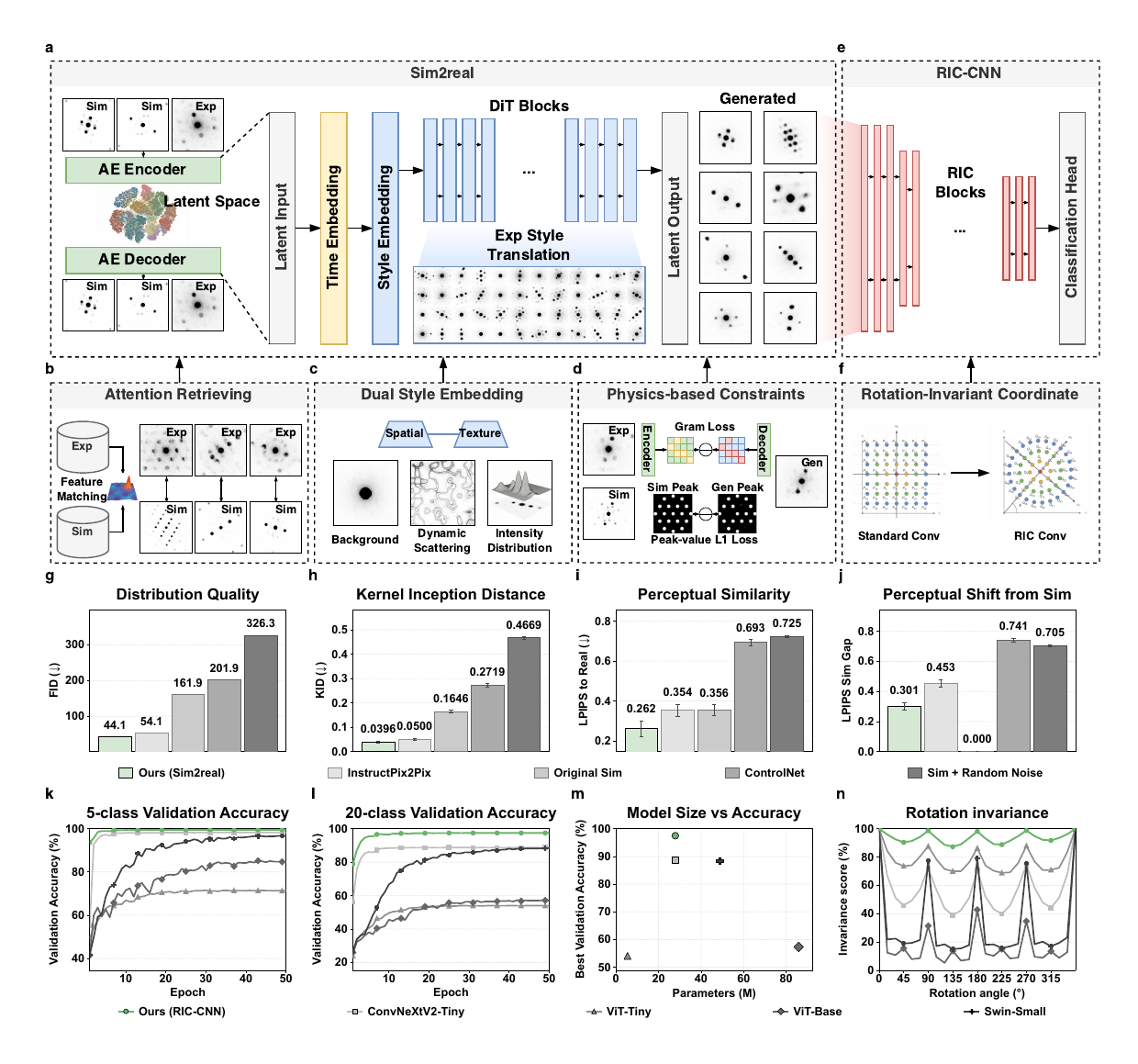}
\caption{\textbf{Architecture and independent benchmarking of the Sim2real and RIC-CNN modules.}
\textbf{a}, Schematic of the Sim2real module, which maps simulated DPs to experimental-style outputs in a shared latent space.
\textbf{b}, Attention-based retrieval of experimentally matched references for Sim2real conditioning.
\textbf{c}, Dual style embedding capturing complementary spatial and textural characteristics of experimental diffraction.
\textbf{d}, Physics-based constraints for preserving experimental appearance while maintaining Bragg peak geometry during translation.
\textbf{e}, Schematic of the RIC-CNN classifier for final component identification.
\textbf{f}, Illustration of rotation-invariant coordinate convolution.
\textbf{g}, Comparison of distribution quality measured by Fr\'{e}chet inception distance (FID).
\textbf{h}, Comparison of distribution quality measured by kernel inception distance (KID).
\textbf{i}, Comparison of perceptual similarity to real experimental patterns measured by LPIPS.
\textbf{j}, Comparison of perceptual shift from the original simulated patterns measured by LPIPS.
\textbf{k}, Validation accuracy curves for the five-class task.
\textbf{l}, Validation accuracy curves for the twenty-class task.
\textbf{m}, Model size versus best validation accuracy.
\textbf{n}, Rotation invariance analysis across input rotation angles.}
\label{fig:modules_benchmarking}
\end{figure*}

It is worth to emphasize that before entering the neural network workflow, experimental DPs are preprocessed to correct tractable experimental variations. This preprocessing includes beam-center and elliptical-distortion corrections\cite{ref33}, together with denoising\cite{ref27}, to reduce systematic geometric artifacts and experimental noise while preserving the diffraction features required for downstream domain alignment and multi-component classification (see Supplementary Fig.~2). After these corrections, the DPs are examined for signals beyond the transmitted central disk. Probe positions showing neither off-center diffraction spots nor a diffuse amorphous ring is treated as empty, while patterns containing only an amorphous ring and no discrete Bragg reflections are labeled as purely amorphous. These patterns are removed before candidate-component analysis (see Supplementary Fig.~3). Because this screening is based on directly observable diffraction features, rather than a learned model, the neural network can focus on the more difficult task of identifying overlapping crystalline components in heterogeneous regions.

Next, we illustrate the two modules in 4DMulti in details. The first module of 4DMulti, termed Sim2real, is designed to address the first challenge outlined above related to the domain gap between idealized diffraction simulations and experimental data (Fig.~\ref{fig:modules_benchmarking}a). In experimental 4D-STEM, DPs essentially are the scattering information arise from not only the underlying crystallographic structure, but also the dynamical scattering, orientational overlap, reduced crystallinity, crystal defects, thickness variation, and low-dose acquisition. Taken together, these effects can weaken, broaden, or obscure high-frequency Bragg features that are critical for reliable component identification. By contrast, multi-slice simulations remain comparatively idealized and merely encode the diffraction response of the presumable structure under controlled conditions. In our experiments, we found that classifiers trained directly on simulated DPs are poorly generalized to experimental data. 4DMulti therefore introduces a dedicated Sim2real module that translates labeled simulated DPs into experimental-style counterparts. We formulated the translation as a physics-preserving domain alignment problem rather than generic image stylization. Although experimental diffraction appearance can vary across material systems, microscopes, specimen thicknesses, and acquisition conditions, the Bragg configuration that encodes structural identity must remain unchanged. The Sim2real module is therefore designed to include experimental appearance statistics while preserving the diffraction geometry required for downstream classification.

In addition, because simulated and experimental DPs are not naturally paired, 4DMulti constructs pseudo-paired guidance through retrieval and a shared latent diffraction representation as shown in Fig.~\ref{fig:modules_benchmarking}b. For each simulated input, the model uses attention-based matching in a global feature space to retrieve the most similar experimental reference.

The retrieved examples thus guide the translation toward the target experimental domain by capturing its specific imaging characteristics, without introducing arbitrary peak displacements. To further constrain the process within a physical framework, both simulated and experimental DPs are encoded into a shared latent space using a diffraction-specialized autoencoder.

A retrieval-conditioned latent Diffusion Transformer then performs the translation under explicit physics-based constraints. Within this shared latent space, dual style embeddings encode complementary spatial and textural characteristics of experimental diffraction, including background statistics, dynamical scattering, and intensity distributions (Fig.~\ref{fig:modules_benchmarking}d). The translation is further constrained by Gram loss and peak-value L1 loss, which align higher-order appearance statistics with the experimental domain while preserving Bragg peak geometry (Fig.~\ref{fig:modules_benchmarking}e). In this way, the Sim2real module achieves domain alignment while maintaining the structural fidelity of the simulated DPs.

The second module of 4DMulti performs final component recognition using a rotation-invariant classifier, termed RIC-CNN (Fig.~\ref{fig:modules_benchmarking}e)\cite{ref32}. This module is designed to handle the strong rotational variation commonly observed in DPs, which can be difficult for standard vision models. Once the patterns are recentered, rotating DPs usually indicates a change in local crystal orientation rather than a different material component. The classifier must therefore remain insensitive to orientation while retaining the Bragg-pattern features needed to distinguish crystallographic components.

Specifically, RIC-CNN replaces standard convolutions with rotation-invariant coordinate convolutions defined with respect to the diffraction centre (Fig.~\ref{fig:modules_benchmarking}f)\cite{ref32,ref34}. Rather than relying on a fixed Cartesian sampling grid, these operations are formulated in a centre-referenced coordinate system, allowing equivalent diffraction structures to remain comparable under arbitrary in-plane rotation. By stacking such rotation-aware operations, RIC-CNN learns feature representations that are robust to rotational variation while preserving structurally informative diffraction geometry.

Together, these two modules define the core logic of 4DMulti. Sim2real aligns simulated DPs with the experimental domain while preserving diffraction structure, whereas RIC-CNN enables robust component recognition despite rotational variation. Their integration allows direct, automated, and spatially resolved identification of local crystallographic components in complex 4D-STEM datasets.

\subsection*{Independent benchmarking of the Sim2real and RIC-CNN modules}

To benchmark the two core modules of 4DMulti, we performed separate evaluations of Sim2real and RIC-CNN. For Sim2real, domain alignment was quantified using Fr\'echet inception distance (FID), kernel inception distance (KID), and LPIPS against experimental diffraction data in Fig.~\ref{fig:modules_benchmarking}g--i (see Supplementary Note for metric definitions and interpretation). To assess whether the translation preserved the crystallographic information in the original simulation without introducing spurious diffraction features, we further measured LPIPS between the translated outputs and the simulated inputs (Fig.~\ref{fig:modules_benchmarking}j). We compared our method against InstructPix2Pix and ControlNet in addition to the unmodified simulated patterns (Origin) and a simple noise-augmented baseline (+Noise) as references. Overall, Sim2real outperformed other models on all three metrics of experimental-domain alignment, while ranking second to the simulated input in source-consistency comparison. These results confirm that Sim2real narrows the simulation--experiment gap while preserving the crystallographic content of the DPs.

\begin{figure*}[!htp]
  \centering
  \includegraphics[width=\textwidth]{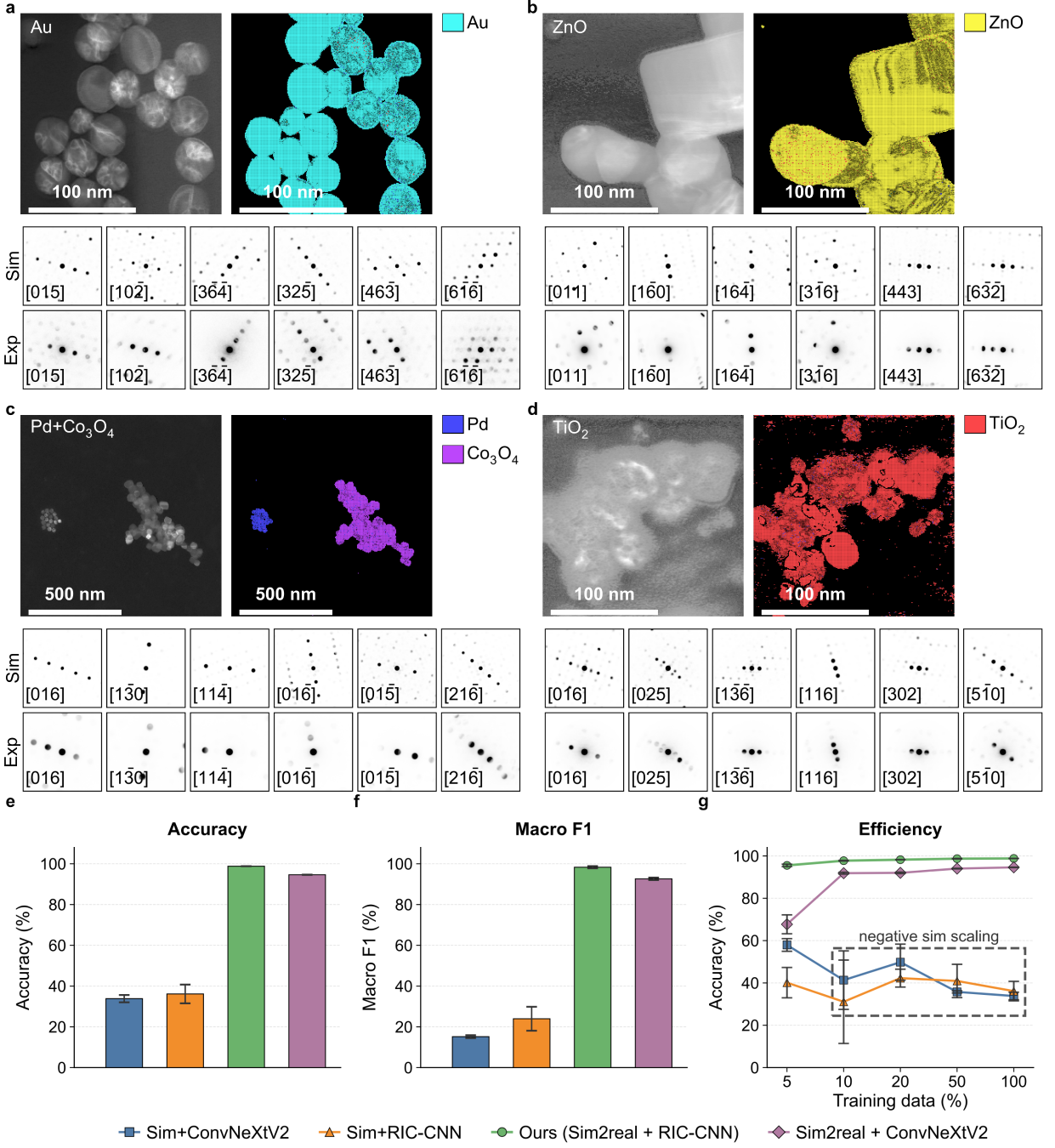}
\caption{\textbf{Experimental benchmarking and ablation analysis of 4DMulti.}
\textbf{a--d}, Representative 4DMulti results for Au, ZnO, Pd + Co$_3$O$_4$, and TiO$_2$, respectively, showing experimental real-space images, predicted component maps, and matched simulated and experimental DPs.
\textbf{e}, Accuracy comparison for 4DMulti and ablation baselines.
\textbf{f}, Macro F1 comparison for 4DMulti and ablation baselines.
\textbf{g}, Training-data efficiency comparison across different workflows.
In \textbf{e--g}, bars and data points represent the mean of five runs ($n=5$), and error bars indicate $\pm$ one standard deviation (SD).
Scale bars: 100~nm in the real-space images and component maps in \textbf{a}, \textbf{b}, and \textbf{d}; 500~nm in \textbf{c}.}
\label{fig:nanoparticle_benchmark}
\end{figure*}

In addition, RIC-CNN was benchmarked against representative convolutional and transformer-based architectures on both five-class and twenty-class classification tasks using the Sim2real-aligned diffraction dataset. The validation curves show that RIC-CNN consistently outperformed the baseline models in both settings, particularly in twenty-class regime in Fig.~\ref{fig:modules_benchmarking}k,l). We further compared best validation accuracy against parameter count on the twenty-class task and found that RIC-CNN exceeds several substantially larger baseline models, indicating an advantage in parameter efficiency (Fig.~\ref{fig:modules_benchmarking}m). Finally, we evaluated rotation robustness by rotating input DPs over a full angular range and measuring the consistency of the resulting latent representations relative to the unrotated input. RIC-CNN exhibited the strongest rotation invariance among the compared models (Fig.~\ref{fig:modules_benchmarking}n). Taken together, these results validate the effectiveness of the two core modules underlying 4DMulti.

\subsection*{Accurate multi-component classification of experimental 4D-STEM DPs}

We next validated 4DMulti as an end-to-end workflow on 4D-STEM dataset acquired from various nanoparticles comprising Au, ZnO, TiO$_2$, Pd, and Co$_3$O$_4$. This dataset serves as a standard test for crystallographic classification, as it represents a diverse range of crystal structures and varying nanoparticle morphologies under realistic conditions. On this experimental benchmark, 4DMulti achieved a classification accuracy of 98.82\%. Notably, the model occasionally fails to correctly classify highly symmetric DPs, representing a limitation in orientation-agnostic mapping. Nonetheless, the representative results, illustrated in Fig.~\ref{fig:nanoparticle_benchmark}a--d, show precise agreement between the translated experimental DPs and their simulated crystallographic templates along with accurate identification of the nanoparticle boundaries. Further matched Sim--Gen--Exp examples are provided in Supplementary Figs.~4--7.

To determine whether these performance gains require our integrated architecture, we compared the full 4DMulti workflow against standard computer vision backbones and individual ablation baselines under identical training conditions. Specifically, we compared the full framework against a conventional backbone trained directly on simulated DPs, an RIC-CNN classifier without Sim2real alignment, and a Sim2real-enhanced backbone lacking RIC-CNN. The complete 4DMulti model achieved the highest accuracy and macro F1 score on the five-class benchmark shown in Fig.~\ref{fig:nanoparticle_benchmark}e,f, demonstrating that neither domain alignment nor rotation-invariant recognition alone is sufficient. A robust multi-component classification requires the synergistic integration of both modules within a unified workflow.

Beyond peak performance, we examined how each model responded to increasing amounts of simulated training data in Fig.~\ref{fig:nanoparticle_benchmark}g. The full Sim2real + RIC-CNN workflow achieved the highest experimental-domain accuracy at every training fraction, with the clearest advantage in the low-data regime. Trained on the same Sim2real-aligned dataset, RIC-CNN substantially outperformed ConvNeXtV2 when only 5\% of the training data were available and remained more accurate as the dataset increased, although the gap gradually narrowed. Because the domain-alignment component was identical in this comparison, the result points to the classifier architecture as the main source of the difference. In particular, the rotation-invariant design of RIC-CNN is better suited to the rotational variability of DPs, enabling the network to learn crystallographic features from fewer training examples than a general-purpose vision backbone.

A different trend emerged when the classifiers were trained directly on unaligned simulated data. Neither model showed a consistent improvement in experimental-domain accuracy as more simulated data were added. Instead, their performance fluctuated and declined at larger training fractions, indicating negative scaling under simulation-to-experiment domain shift. This pattern is consistent with the models becoming increasingly dependent on simulation-specific cues that do not transfer reliably to experimental DPs. Although the result does not by itself prove simulation-domain overfitting, it shows that increasing the amount of unaligned simulated data is not sufficient to improve cross-domain generalization and may even reduce it. Taken together, these findings indicate that Sim2real alignment is necessary for additional simulated data to translate into better experimental performance, while the diffraction-specific design of RIC-CNN provides a further advantage, particularly when training data are limited.

\subsection*{Resolving a complex superconducting interface with 4DMulti}

\begin{figure*}[!htp]
  \centering
  \includegraphics[width=\textwidth]{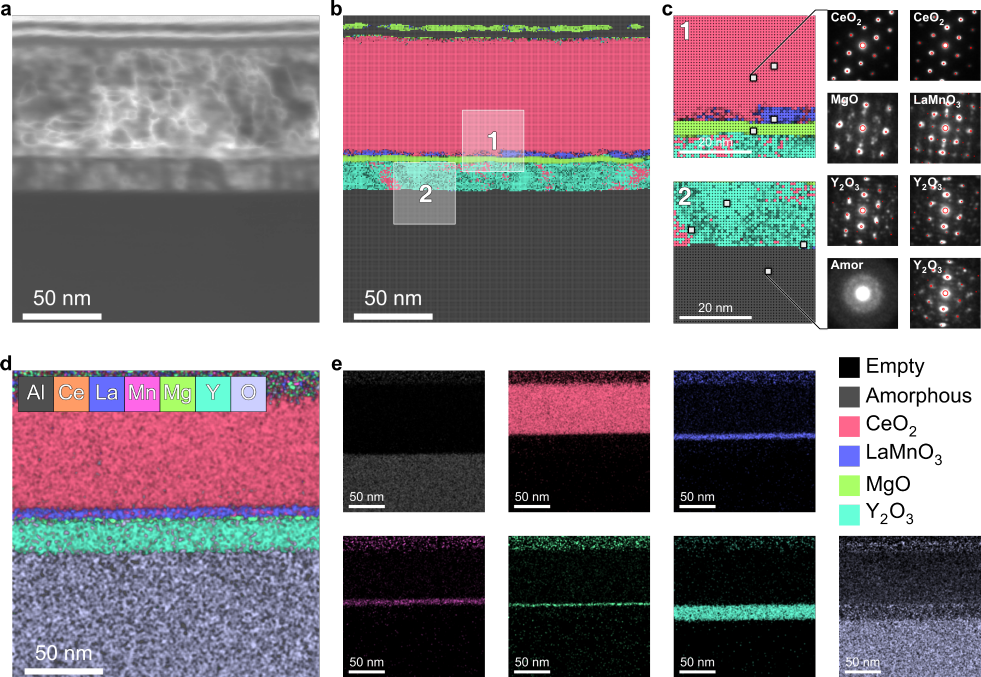}
\caption{\textbf{Phase mapping of a multilayer oxide heterostructure in a superconducting device using 4DMulti.}
\textbf{a}, Cross-sectional bright-field STEM image of the multilayer oxide thin film.
\textbf{b}, 4DMulti-predicted spatial phase classification map, resolving layers of CeO$_2$ (pink), LaMnO$_3$ (blue), MgO (green), and Y$_2$O$_3$ (cyan), along with amorphous (grey) and vacuum/empty (black) regions. White boxes (1 and 2) mark regions chosen for closer inspection.
\textbf{c}, Magnified views of regions 1 and 2 (left) with representative experimental DPs (right) sampled from single-pixel positions (white squares). Red circles and overlays indicate indexed Bragg reflections corresponding to CeO$_2$, MgO, LaMnO$_3$, and Y$_2$O$_3$, while diffuse scattering identifies the amorphous region. These results highlight nanoscale spatial resolution and the detection of localized CeO$_2$ nanodomains embedded within the Y$_2$O$_3$ layer.
\textbf{d}, Composite energy-dispersive X-ray spectroscopy (EDX) elemental map combining signals from Al, Ce, La, Mn, Mg, Y, and O.
\textbf{e}, Individual EDX elemental maps of each corresponding element, corroborating the structural phase boundaries and layer assignments identified by 4DMulti.
Scale bars: 50~nm in \textbf{a}, \textbf{b}, \textbf{d}, and \textbf{e}; 20~nm in \textbf{c}.}
\label{fig:superconducting_interface}
\end{figure*}

We next applied 4DMulti to a multilayer oxide heterostructure in a commercial superconducting device. Compared with the nanoparticle benchmark, the focused ion beam (FIB)-prepared cross-section contains a more heterogeneous arrangement of crystalline and amorphous regions at the nanoscale. As shown in Fig.~\ref{fig:superconducting_interface}, 4DMulti identifies an upper CeO$_2$ layer, ultrathin LaMnO$_3$/MgO interfacial regions, a buried Y$_2$O$_3$ layer, and a lower amorphous Al-containing layer. The phase map separates the closely spaced buffer layers and reveals local departures from the nominal layered structure.

Closer inspection reveals localized CeO$_2$ nanodomains within the Y$_2$O$_3$ layers in Fig.~\ref{fig:superconducting_interface}c, indicating that its distribution is not confined to the upper layer. These domains are not clearly resolved in the corresponding STEM image or energy-dispersive X-ray spectroscopy (EDX) maps, illustrating the additional structural information provided by diffraction-based identification. 

The mapped layers agree with the expected device structure and the layer boundaries visible in the STEM image. Representative experimental DPs match simulations of the assigned crystal structures, while the lower Al-containing layer shows diffuse scattering consistent with an amorphous structure. EDX mapping further supports the overall layer assignments (Fig.~\ref{fig:superconducting_interface}d,e): Ce is concentrated in the upper layer; La, Mn, and Mg lie near the buffer layer; Y is concentrated in the seed layer; and Al and O dominate the lower region. Together, our observations distinguish the crystalline oxide layers from the amorphous Al-rich region and identify local phase variations that are not apparent from elemental contrast alone.

These results extend the evaluation of 4DMulti from nanoparticle references to a heterogeneous interface in an engineering material. It is worth to point out that although the principal layers show spatially coherent phase assignments, transition regions exhibit local variations accompanied by weak and overlapping Bragg reflections in Fig.~\ref{fig:superconducting_interface}b,c. A single phase label cannot fully represent such diffraction signals, which may contain contributions from multiple structures along the beam direction. This motivates complementing the phase maps with a spatially resolved measure of assignment ambiguity.

\subsection*{Predictive uncertainty reveals local interfacial structural complexity}

In the superconducting heterostructure, DPs from layer interiors are generally dominated by a single phase. In more complex heterogeneous interfaces, overlapping DPs from multiple phases along the beam direction are frequently observed, which poses a major challenge for resolving individual components. Interestingly, 4DMulti provides a mathematically natural route to quantify this complexity because the classifier outputs a probability distribution over the physically plausible candidate phases. Rather than treating this distribution solely as a confidence score, we interpret its spread as predictive uncertainty within the candidate structural space. To project this uncertainty into a spatially resolved descriptor of interfacial heterogeneity, we define diffraction-inferred structural complexity (DISC) as the normalized predictive entropy of the classifier output.

The rationale for defining such a quantity from the network output is rooted in the structure of the 4DMulti pipeline. For each DP, the RIC-CNN classifier outputs a class-probability vector:

\begin{equation}
\mathbf{p}=(p_1,p_2,\ldots,p_K),
\label{eq:class_probability_vector}
\end{equation}

where $K$ is the number of candidate components and $p_k$ denotes the predicted probability assigned to the $k$th component. The largest entry in this vector determines the primary component assignment, but the full distribution contains additional information about how decisively the diffraction signal favors that assignment over competing alternatives. The DISC is thereby defined as:

\begin{equation}
\mathrm{DISC}=-\frac{\sum_{k=1}^{K}p_k\log p_k}{\log K},
\label{eq:disc}
\end{equation}

Under this formulation, DISC approaches 0 when the classifier assigns most of the predicted probability to a single component, indicating a decisive structural assignment. Conversely, DISC approaches 1 when the predicted probability is distributed more evenly across multiple plausible components, indicating greater assignment uncertainty within the candidate structural space. The distribution of DISC can be calculated across the entire dataset, reflecting how uniquely the measured diffraction signal supports the principal structural assignment. Therefore, this parameter provides a quantitative and spatially resolved description of local interfacial heterogeneity derived directly from predictive uncertainty.

We emphasize that The candidate phases are specified by crystal structures, and the classifier is trained on simulation-derived patterns translated into the experimental domain. DISC therefore describes ambiguity within this selected structural library. A pattern dominated by one recognizable phase may yield low DISC, whereas overlapping or perturbed reflections may produce competing assignments and higher DISC. However, DISC does not measure phase fractions or structural disorder directly. Its value also depends on experimental noise, domain mismatch, similarities among candidate structures, and probability calibration. A low value indicates a decisive prediction, but does not establish that the assignment is correct, particularly when the true phase is absent from the library.

This entropy-based descriptor is validated by nanoparticle references and independently corroborated on the superconducting interface via prior structural knowledge, DP matching, and EDX mapping. Based on these results, we propose that DISC can serve as a standard parameter in 4D-STEM structural identification. Visualizing the spatial distribution of DISC alongside classification maps ensures the fidelity of the phase assignments while revealing local variations that are invisible to hard, discrete classification alone. Incorporating this parameter enables 4DMulti to offer a quantitative, spatially resolved description of structural complexity across diverse material systems, moving automated microscopy closer to high-throughput industrial quality control.

\subsection*{Generalization across diverse interfacial material systems}

To extend the generality of 4DMulti as a unified framework for complex interfacial analysis, we further applied it to material systems with fundamentally different chemistries, crystallographic motifs, and degrees of structural heterogeneity. Beyond the superconducting interface, we examined two representative cases: a corroded Mg-alloy interface and a degraded sulfur-based solid-electrolyte interface. Together, these systems encompass markedly different crystal structures and interfacial morphologies, providing a stringent test of the framework's transferability across material classes. For each system, we generated phase maps and corresponding confidence and DISC maps to examine the phase distribution and identify regions with ambiguous assignments.

\begin{figure*}[!htp]
    \centering
    \includegraphics[
        width=0.9\textwidth,
        keepaspectratio
    ]{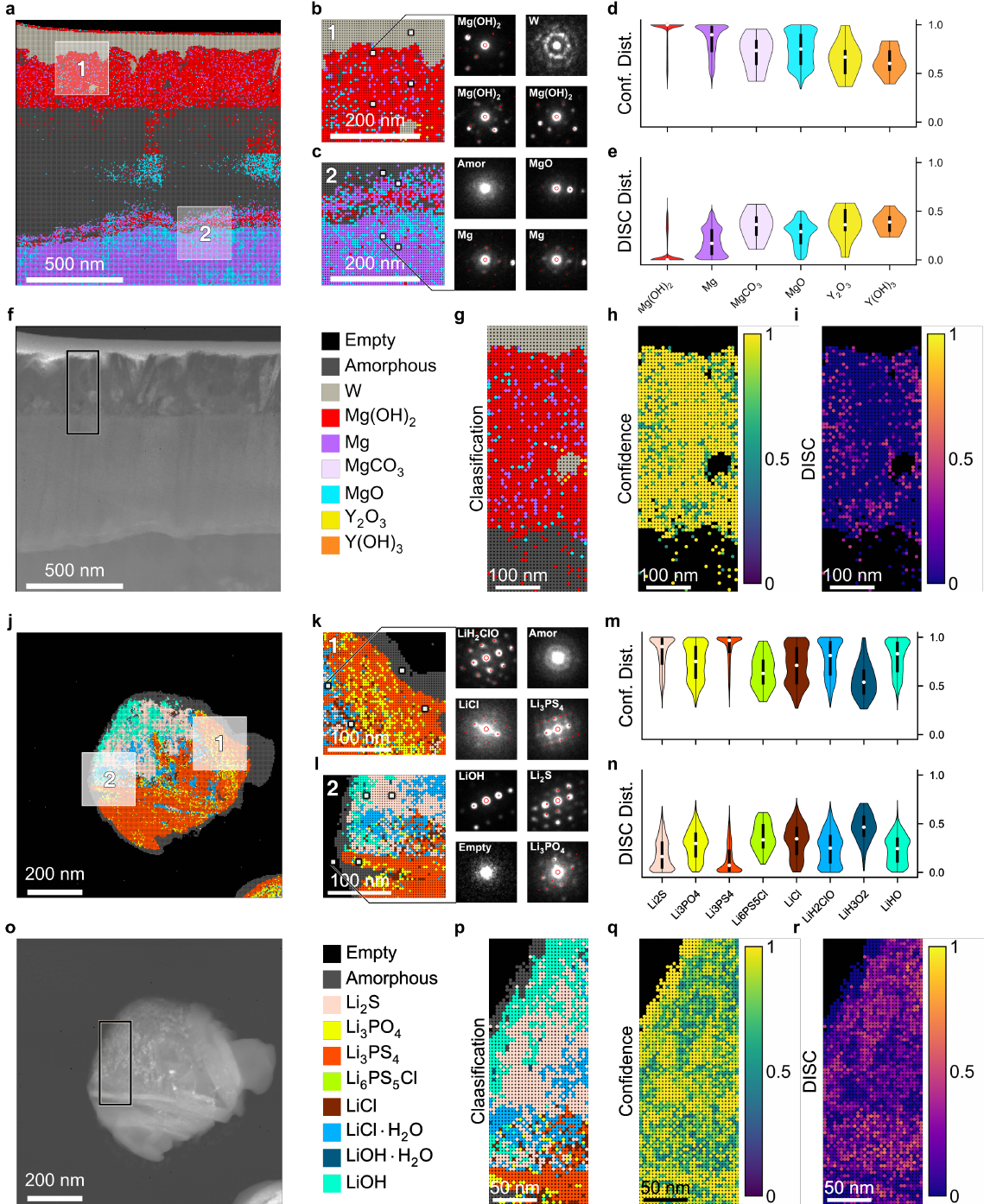}
\caption{\textbf{Phase mapping and diffraction-inferred structural complexity at corrosion and degradation interfaces.}
\textbf{a--i}, Corroded Mg-alloy interface.
\textbf{a}, 4DMulti phase map, with boxes 1 and 2 marking regions selected for closer inspection.
\textbf{b,c}, Enlarged views of regions 1 and 2, respectively, with experimental DPs from the marked scan positions. Red circles indicate indexed Bragg reflections.
\textbf{d}, Violin plots of predictive confidence, defined as the highest predicted class probability, grouped by assigned phase. 
\textbf{e}, Corresponding distributions of diffraction-inferred structural complexity (DISC), defined as normalized predictive entropy. 
\textbf{f}, STEM image with the black rectangle marking the region shown in \textbf{g--i}.
\textbf{g--i},Phase map \textbf{g}, predictive confidence map \textbf{h}, and DISC map \textbf{i}.
\textbf{j--r}, Moisture-degraded Li$_6$PS$_5$Cl solid electrolyte.
\textbf{j}, 4DMulti phase map, with boxes 1 and 2 marking regions selected for closer inspection.
\textbf{k,l}, Enlarged views of regions 1 and 2, respectively, with experimental DPs from the marked scan positions. Red circles indicate indexed Bragg reflections.
\textbf{m,n}, Predictive confidence \textbf{m} and DISC \textbf{n} distributions grouped by assigned phase.
\textbf{o}, STEM image, with the black rectangle marking the region shown in \textbf{p--r}.
\textbf{p--r},Phase map \textbf{p}, predictive confidence map \textbf{q}, and DISC map \textbf{r}. Phase colours are defined in the corresponding legends. Amor and Empty denote amorphous and empty regions, respectively. Higher DISC indicates greater assignment ambiguity within the candidate phase library. Scale bars: 500~nm in \textbf{a,f}; 200~nm in \textbf{b,c,j,o}; 100~nm in \textbf{g--i,k,l}; 50~nm in \textbf{p--r}.
}
\label{fig:interfacial_generalization}
\end{figure*}

For the Mg-alloy case, we examined an Mg--11 wt.\% Y--1 wt.\% Al alloy after 1.5 h of exposure to a 3.5 wt.\% NaCl solution in Fig.~\ref{fig:interfacial_generalization}a--i. From the surface inward, the cross-section contains a protective W layer deposited during FIB preparation, a corrosion-product layer, an amorphous interlayer, and the alloy substrate. The phase map identifies crystalline Mg(OH)$_2$ and MgO in the corrosion products, with Mg(OH)$_2$ occupying much of the outer layer. Representative DPs support these assignments in Fig.~\ref{fig:interfacial_generalization}b. The coexistence of crystalline products and an amorphous interlayer shows that the corrosion region is structurally heterogeneous through its thickness.

Closer to the substrate, MgO is also identified among regions assigned to metallic Mg as shown Fig.~\ref{fig:interfacial_generalization}c. This distribution suggests that oxide formation is not confined to the outer corrosion layer. The present measurements cannot distinguish oxidation during NaCl exposure from that introduced during specimen preparation or subsequent handling. Nevertheless, the map locates oxide-bearing regions within the predominantly metallic area that would be difficult to distinguish from morphology alone.

We next examined a moisture-degraded Li$_6$PS$_5$Cl argyrodite solid electrolyte in Fig.~\ref{fig:interfacial_generalization}j--r. Within the selected candidate library, 4DMulti assigns crystalline regions to residual Li$_6$PS$_5$Cl and seven additional phases: Li$_2$S, Li$_3$PO$_4$, Li$_3$PS$_4$, LiCl, LiCl$\cdot$H$_2$O, LiOH$\cdot$H$_2$O, and LiOH. Representative indexed patterns support local assignments in Fig.~\ref{fig:interfacial_generalization}k,l. The maps show an uneven distribution of these phases across the particle, with residual electrolyte present alongside several degradation-related products. This spatial coexistence is consistent with incomplete and heterogeneous degradation.

We want to point out that the spatial distribution of DISC reveals a consistent qualitative trend: bulk regions far from interfaces exhibit low structural complexity, whereas phase boundaries, degradation fronts, and transition zones display significantly elevated complexity in Fig.~\ref{fig:interfacial_generalization}i,r. This observation is reasonable as single-phase bulk regions yield stable, highly defined diffraction signatures, while transition layers contain overlapping or locally perturbed Bragg reflections that broaden the classifier's probability distribution. The recurrence of this spatial trend across diverse material systems indicates that DISC is not an artifact of a specific dataset, but a generalizable physical descriptor of local structural ambiguity. By jointly mapping dominant phases and quantifying structural complexity, the framework captures the spatial extent and internal disorder of interfacial transition zones where discrete classification labels are insufficient.

These applications show that 4DMulti can map candidate phases in both a layered corrosion interface and a heterogeneous degraded particle. In the alloy, it distinguishes crystalline corrosion products, an amorphous interlayer, and oxide-bearing regions near the substrate. In the electrolyte, it maps residual material alongside several degradation-related phases and amorphous regions. The accompanying DISC maps identify where these assignments are less decisive and where additional characterization would be most useful.

\section*{Discussion}

4DMulti links simulation-derived crystallographic labels to experimental 4D-STEM data to produce spatially resolved phase maps. Given a suitable candidate phase library, it assigns patterns across a scan without manual interpretation of each measurement. The nanoparticle benchmark and the three interfacial applications demonstrate its use in mapping phase distributions across materials with different compositions and morphologies.

The benchmarking results indicate that domain alignment and rotation-invariant classification address complementary sources of error. Sim2real reduces differences between simulated and experimental DPs, while RIC-CNN accommodates in-plane rotations without rotational data augmentation. Their combination achieves the highest experimental classification accuracy among the tested methods, with a particularly clear advantage when training data are limited. Increasing the amount of unaligned simulated data, by contrast, does not consistently improve experimental performance. These results suggest that the usefulness of simulated training data depends on both their correspondence to experimental measurements and the classifier's treatment of diffraction geometry.

Each phase label represents the highest-probability candidate within the selected library. It does not provide a complete decomposition of contributions along the beam direction. DISC supplements these labels by describing the spread of predicted probabilities. Its elevated values in parts of the interfacial transition regions identify positions where phase assignments are less decisive than in the surrounding layer interiors.

The interpretation of DISC depends on the candidate library and probability calibration. High values cannot distinguish phase coexistence or disorder from noise, domain mismatch, or similar diffraction signatures among candidates. Low values likewise do not establish correctness when the relevant phase is absent or the model is overconfident. DISC therefore measures assignment ambiguity rather than phase fractions or structural disorder directly, and comparisons across datasets require comparable candidate libraries and calibration. Although diffuse scattering allows amorphous regions to be separated from crystalline regions, the present workflow does not distinguish different forms of short-range order. Strongly overlapping crystalline contributions also remain difficult to resolve. Models that explicitly account for mixtures, representations of diffuse scattering, and detection of phases outside the candidate library are possible extensions.

In summary, 4DMulti offers a route towards automated phase identification at increasingly complex material interfaces. As the crystal-structure library expands, the framework could identify a wider range of constituent phases without requiring prior knowledge of their spatial arrangement. The long-term goal is to analyze interfaces of unknown composition and architecture, provided their constituent structures are represented in the library and distinguishable in the measured diffraction data. Combined with methods for resolving phase overlap and detecting structures outside the library, this approach could extend 4D-STEM to automated laboratory workflows and industrial applications, including failure analysis, process development, and quality assessment.

\section*{Data availability}

The datasets generated and analyzed during the current study are not publicly available due to the internal data sharing policies of the participating institutions, but are available from the corresponding author on reasonable request.

\section*{Code availability}

The code will be released via a public GitHub repository upon publication to support reproducibility and community use.

\bibliography{4DMulti_refs}

\section*{Acknowledgments}

The authors thank the Instrumental Analysis Center of Shanghai Jiao Tong University for the support.

\section*{Author contributions}

H.Z., Z.M. and Y.X. conceived and designed the study. X.H., Y.G., and S.C. prepared the specimens and performed the experimental 4D-STEM acquisitions. H.Z., M.L., and A.Y. conducted and developed the model. H.Z. and Y.X. drafted the paper. X.Z. and Y.X. supervised the project. All authors reviewed and approved the final paper.

\section*{Funding}

This work was financially supported by National Natural Science Foundation of China grant 12474186.

\section*{Competing interests}

Authors declare that they have no competing interests.

\end{document}